\documentclass[letterpaper,twocolumn,prl,aps,superscriptaddress,amsmath,amssymb,floatfix]{revtex4-2}
\usepackage{mathptmx}
\usepackage[latin9]{inputenc}
\usepackage{color}
\usepackage{verbatim}
\usepackage{siunitx}
\usepackage{upgreek}
\usepackage{amssymb,amsthm,amsmath,amssymb,bbm}
\usepackage{adjustbox}
\usepackage{esint}
\usepackage[percent]{overpic}
\usepackage{graphicx,float,subfigure}
\usepackage{dcolumn}
\usepackage{bm,braket,color}

\usepackage[unicode=true,
bookmarks=true,bookmarksnumbered=false,bookmarksopen=false,
breaklinks=false,pdfborder={0 0 1},backref=false,colorlinks=true]
{hyperref}
\hypersetup{
	linkcolor=magenta,urlcolor=blue,citecolor=blue,pdfstartview={FitH},hyperfootnotes=false}
\makeatletter

\usepackage{textcomp}
\usepackage{epstopdf}

\pdfpageheight\paperheight
\pdfpagewidth\paperwidth

\@ifundefined{textcolor}{}{%
	\definecolor{BLACK}{gray}{0}
	\definecolor{WHITE}{gray}{1}
	\definecolor{RED}{rgb}{1,0,0}
	\definecolor{GREEN}{rgb}{0,1,0}
	\definecolor{BLUE}{rgb}{0,0,1}
	\definecolor{CYAN}{cmyk}{1,0,0,0}
	\definecolor{MAGENTA}{cmyk}{0,1,0,0}
	\definecolor{YELLOW}{cmyk}{0,0,1,0}
}

\usepackage{xcolor}
\usepackage{soul}
\definecolor{blue}{rgb}{0,0,1}
\definecolor{red}{rgb}{1,0,0}
\definecolor{green}{rgb}{0,1,0}

\DeclareMathAlphabet{\mathcal}{OMS}{cmsy}{m}{n}
\DeclareSymbolFont{largesymbols}{OMX}{cmex}{m}{n}

\usepackage{soul}

\begin{document}

\affiliation{Laboratory of Quantum Information, University of Science and Technology of China, Hefei 230026, China}
\affiliation{CAS Center for Excellence in Quantum Information and Quantum Physics, University of Science and Technology of China, Hefei, Anhui 230026, China}
\affiliation{Key Laboratory of Quantum Optics and Quantum Optics Devices, Institute of Opto-Electronics, Shanxi University, Taiyuan 030006, China}
\affiliation{Collaborative Innovation Centre of Extreme Optics, Shanxi University, Taiyuan 030006, China}
\affiliation{Hangzhou Biaozhang Electronic Technology Co., Ltd., Hangzhou 311121, China}
\affiliation{International Quantum Academy, Shenzhen,518048, China}
\affiliation{Hefei National Laboratory, University of Science and Technology of China, Hefei 230088, China}

\title{Low-latency FPGA-based electronic control system for fast preparation of defect-free atom arrays}

\author{Ya-Dong Hu}
\thanks{These authors contributed equally to this work.}
\affiliation{Laboratory of Quantum Information, University of Science and Technology of China, Hefei 230026, China}
\affiliation{CAS Center for Excellence in Quantum Information and Quantum Physics, University of Science and Technology of China, Hefei, Anhui 230026, China}

\author{Dong-Qi Ma}
\thanks{These authors contributed equally to this work.}
\affiliation{Laboratory of Quantum Information, University of Science and Technology of China, Hefei 230026, China}
\affiliation{CAS Center for Excellence in Quantum Information and Quantum Physics, University of Science and Technology of China, Hefei, Anhui 230026, China}

\author{Tian-Yang Zhang}
\affiliation{Laboratory of Quantum Information, University of Science and Technology of China, Hefei 230026, China}
\affiliation{CAS Center for Excellence in Quantum Information and Quantum Physics, University of Science and Technology of China, Hefei, Anhui 230026, China}

\author{Liang Chen}
\affiliation{Laboratory of Quantum Information, University of Science and Technology of China, Hefei 230026, China}
\affiliation{CAS Center for Excellence in Quantum Information and Quantum Physics, University of Science and Technology of China, Hefei, Anhui 230026, China}

\author{Yi-Chen Zhang}
\affiliation{Laboratory of Quantum Information, University of Science and Technology of China, Hefei 230026, China}
\affiliation{CAS Center for Excellence in Quantum Information and Quantum Physics, University of Science and Technology of China, Hefei, Anhui 230026, China}

\author{Xiao-Kang Zhong}
\affiliation{Laboratory of Quantum Information, University of Science and Technology of China, Hefei 230026, China}
\affiliation{CAS Center for Excellence in Quantum Information and Quantum Physics, University of Science and Technology of China, Hefei, Anhui 230026, China}

\author{Wen-Yi Zhu}
\affiliation{Laboratory of Quantum Information, University of Science and Technology of China, Hefei 230026, China}
\affiliation{CAS Center for Excellence in Quantum Information and Quantum Physics, University of Science and Technology of China, Hefei, Anhui 230026, China}

\author{Hong-Jie Fan}
\affiliation{Laboratory of Quantum Information, University of Science and Technology of China, Hefei 230026, China}
\affiliation{CAS Center for Excellence in Quantum Information and Quantum Physics, University of Science and Technology of China, Hefei, Anhui 230026, China}

\author{Qing-Xuan Jie}
\affiliation{Laboratory of Quantum Information, University of Science and Technology of China, Hefei 230026, China}
\affiliation{CAS Center for Excellence in Quantum Information and Quantum Physics, University of Science and Technology of China, Hefei, Anhui 230026, China}

\author{Yan-Lei Zhang}
\affiliation{Laboratory of Quantum Information, University of Science and Technology of China, Hefei 230026, China}
\affiliation{CAS Center for Excellence in Quantum Information and Quantum Physics, University of Science and Technology of China, Hefei, Anhui 230026, China}
\affiliation{International Quantum Academy, Shenzhen,518048, China}
\affiliation{Hefei National Laboratory, University of Science and Technology of China, Hefei 230088, China}

\author{Gang Li}
\affiliation{Key Laboratory of Quantum Optics and Quantum Optics Devices, Institute of Opto-Electronics, Shanxi University, Taiyuan 030006, China}
\affiliation{Collaborative Innovation Centre of Extreme Optics, Shanxi University, Taiyuan 030006, China}

\author{Xi-Feng Ren}
\affiliation{Laboratory of Quantum Information, University of Science and Technology of China, Hefei 230026, China}
\affiliation{CAS Center for Excellence in Quantum Information and Quantum Physics, University of Science and Technology of China, Hefei, Anhui 230026, China}
\affiliation{Hefei National Laboratory, University of Science and Technology of China, Hefei 230088, China}

\author{Xu-Liang Zhang}
\affiliation{Hangzhou Biaozhang Electronic Technology Co., Ltd., Hangzhou 311121, China}

\author{Guang-Can Guo}
\affiliation{Laboratory of Quantum Information, University of Science and Technology of China, Hefei 230026, China}
\affiliation{CAS Center for Excellence in Quantum Information and Quantum Physics, University of Science and Technology of China, Hefei, Anhui 230026, China}
\affiliation{Hefei National Laboratory, University of Science and Technology of China, Hefei 230088, China}

\author{Zhu-Bo Wang}
\email{zbwang@ustc.edu.cn}
\affiliation{Laboratory of Quantum Information, University of Science and Technology of China, Hefei 230026, China}
\affiliation{CAS Center for Excellence in Quantum Information and Quantum Physics, University of Science and Technology of China, Hefei, Anhui 230026, China}

\author{Chang-Ling Zou}
\email{clzou321@ustc.edu.cn}
\affiliation{Laboratory of Quantum Information, University of Science and Technology of China, Hefei 230026, China}
\affiliation{CAS Center for Excellence in Quantum Information and Quantum Physics, University of Science and Technology of China, Hefei, Anhui 230026, China}
\affiliation{Hefei National Laboratory, University of Science and Technology of China, Hefei 230088, China}

\begin{abstract}
The scalability of neutral atom quantum computing demands integrated electronic control systems with low latency, modular architecture, and real-time feedback capability. Here, we present an FPGA-based electronic control system that eliminates the PC from the feedback loop, integrating photon counting, real-time decision-making, and waveform generation within a unified PXIe architecture. The system achieves a total feedback latency of $282\,\mathrm{\mu s}$ and is validated in practical experiments by assembling defect-free atom arrays from 24 stochastically loaded optical tweezers. A single-round rearrangement achieves a filling fraction of $\sim96\%$, while feedback-controlled iterative rearrangement over five rounds boosts the success probability for generating a 10-atom defect-free array from $65.7\%$ to $95.4\%$. This system establishes the electronic infrastructure necessary for mid-circuit measurement and real-time quantum error correction on neutral-atom platforms.
\end{abstract}

\maketitle

\section{Introduction}

\indent Neutral atom arrays have emerged as one of the most promising quantum information processing platforms for realizing universal quantum computation~\cite{RN396,bluvsteinLogicalQuantumProcessor2024a,bluvstein2025faulttolerant,chiu2025continuous} and studying many-body physics~\cite{bernienProbingManybodyDynamics2017,ebadiQuantumPhasesMatter2021,evered2025probing,manovitz2025quantum}. With the capability to host thousands of qubits~\cite{chiu2025continuous, manetschTweezerArray61002025, zhuHighefficiencyLoading24002025}, this platform has demonstrated remarkable performance metrics. For instance, the fidelity of global single-qubit gates exceeds $99.99\%$~\cite{sheng2018highfidelity}, and that of local single-qubit gates has reached $99.96\%$~\cite{rozanov2025benchmarking}, while the two-qubit gate fidelity stands at $99.5\%$~\cite{evered2023highfidelity,muniz2025highfidelity,ma2023highfidelitya}. These results demonstrate that neutral atom platforms have surpassed the key thresholds required for quantum error correction (QEC)~\cite{shorFaulttolerantQuantumComputation1996,steane1996multipleparticle}. Based on these physical metrics, the platform of neutral atom arrays has entered the early stage of logical qubit gate operations. Recent experiments have successfully shown the fault-tolerant algorithm on logical processors~\cite{bluvsteinLogicalQuantumProcessor2024a} and demonstrated that the logical error rate can be suppressed by increasing the code distance~\cite{bluvsteinLogicalQuantumProcessor2024a,bluvstein2025faulttolerant}. All of these show that neutral atom systems have great potential for realizing universal quantum computation.

\begin{figure*}[ht]
\centering\includegraphics[width=11cm]{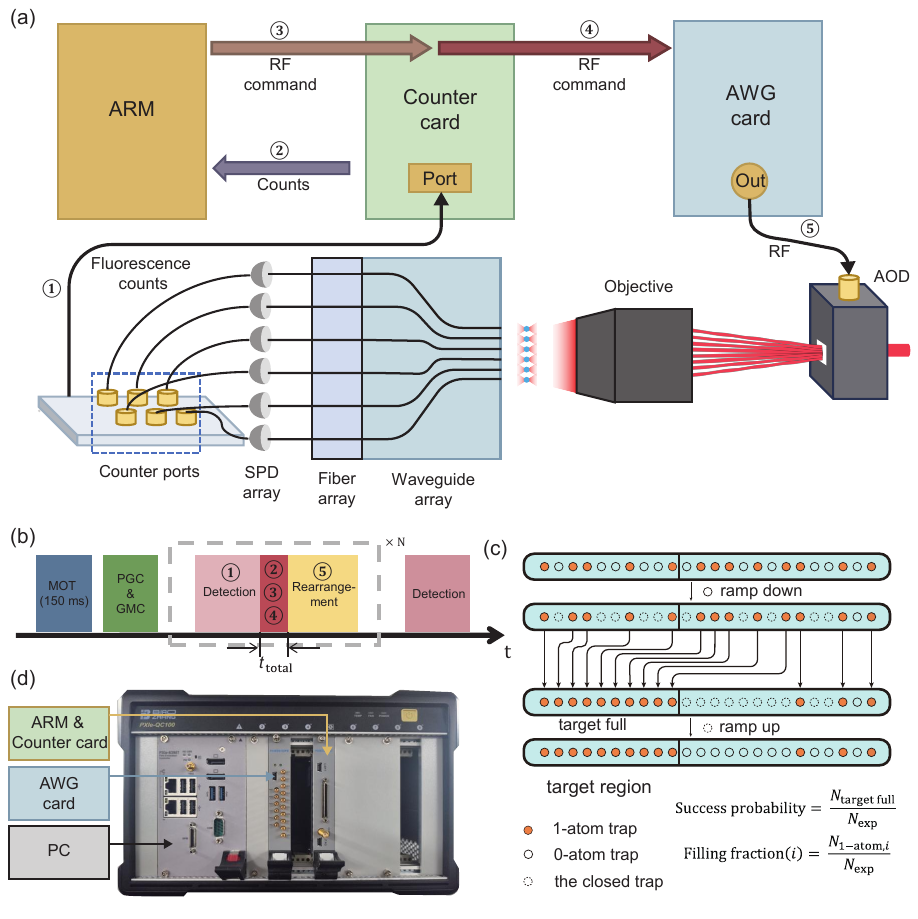}
\caption{ \textbf{Electronic control system and atom rearrangement scheme.} (a) Experimental setup showing the optical path and electronic feedback loop. The one-dimensional (1D) atom array is generated by an acousto-optics deflector (AOD), and fluorescence is collected via optical channel mapping (OCM), fiber arrays (FAs), and SPD arrays (SPDAs). The counter card and ARM processor perform real-time detection and path planning; the arbitrary waveform generator (AWG) card drives the AOD for static tweezers and rearrangement. (b) Timing sequence of the rearrangement process, where $t_{\text{total}}$ is the total electronic latency. (c) Sketch of the 1D rearrangement process with definitions of success probability and filling fraction. The black vertical line divides the array into two regions: the target zone (left), where the defect-free atom array is assembled, and the reservoir zone (right), which stores surplus atoms for replenishing defects in the target zone. (d) Physical implementation of the PXIe-QC100 system with up to 8 slots for counter and AWG cards.}
\label{Fig1}
\end{figure*}

\indent However, despite breakthroughs in gate fidelity at the physical qubit level and system scalability, neutral atom array platforms still face several critical challenges~\cite{awschalom2025challenges} to realize their full potential for achieving fault-tolerant quantum computation~\cite{shorFaulttolerantQuantumComputation1996,steane1996multipleparticle}. One important challenge is that there is no dedicated electronic control system enabling real-time feedback. These real-time feedback capabilities include defect-free atom array assembly, real-time compensation for atom loss, mid-circuit measurement, and real-time QEC. Consequently, current demonstrations of QEC rely on data post-processing~\cite{bluvsteinLogicalQuantumProcessor2024a} due to the lack of a dedicated control system.  The absence of dedicated, scalable, low-latency hardware control systems has become a key bottleneck that prevents neutral atom platforms from realizing practical fault-tolerant quantum processors. Therefore, developing integrated hardware systems for high-speed readout, real-time decoding, and real-time feedback for neutral atom QEC is a critical challenge.

\indent A typical neutral atom quantum circuit comprises initial state preparation, single-qubit gates, two-qubit gates, and readout. The typical timescales are as follows. Initial state preparation requires ${1}-{10}\,\mathrm{\mu s}$, and to ensure preparation efficiency, we always employ 10 times this duration~\cite{chiu2025continuous}. Single-qubit gates implemented via Raman transitions can achieve gate operations in ${\sim2}\,\mathrm{\mu s}$~\cite{wintersperger2023neutral}. Two-qubit gates utilizing Rydberg state interactions require ${\sim1}\,\mathrm{\mu s}$~\cite{levine2019parallel}. Readout operations typically take ${10}-{200}\,\mathrm{\mu s}$~\cite{muzifalconi2025microsecondscale,shea2020submillisecond,chow2023highfidelity}. The total execution time for a complete circuit is on the order of hundreds of microseconds. To enable real-time feedback within this timescale, the PC must be removed from the feedback loop and relegated solely to a configuration role, with all data acquisition, processing, and feedback logic transferred to the FPGA.

\indent Here, we present an FPGA-based electronic control system that eliminates the PC from the feedback loop, integrating photon counting, real-time decision-making, and waveform generation within a unified PXIe architecture. We reduce the total latency of the electronic system to approximately {282(19)}{\,$\mathrm{\mu s}$} during the atom rearrangement, achieving a significant improvement compared to PC-based feedback systems with latencies exceeding {7}\,{ms}~\cite{dadpour2025lowlatencya}.  We employ a volcano architecture employing an optical channel mapping~\cite{ma2025volcano} for parallel readout and the FPGA control system for real-time feedback, and characterize the electronic control system performance by assembling a target array of $N=10$ atoms from 24 traps. While single-round rearrangement yields a success probability of $65.68\%$, the low-latency real-time electronic hardware enables multiple feedback iterations, significantly improving the final success probability to $95.38\%$ for the generation of the defect-free atom array for $N=10$. This low-latency electronic control system provides robust hardware support for high-speed atom assembly and new opportunities for mid-circuit measurements~\cite{grahamMidcircuitMeasurementsSingleSpecies2023,ma2023highfidelitya} and real-time QEC, when combined with fast, non-destructive readout techniques~\cite{chow2023highfidelity,shea2020submillisecond,radnaevUniversalNeutralAtomQuantum2025}.

\section{Experimental system}
\label{sec:setup}

Figure~\ref{Fig1}(a) illustrates the experimental system, a prototype quantum processing unit built around a single-atom array. The system separates into four modules: (i) the atomic array held in vacuum; (ii) a \emph{classical link} that delivers the trapping and control light to the target atoms; (iii) a \emph{quantum link} that collects the single-photon fluorescence emitted by individual atoms; and (iv) a \emph{classical control module} that processes the detection signals, executes the feedback logic, and generates the electronic drive signals returned to the classical link. Here we adopt the volcano architecture~\cite{ma2025volcano}, introducing an optical waveguide array for optical channel mapping (OCM) so that photons emitted by different atoms are routed separately to individual single-photon detectors. The core innovation of this work is the low-latency control module: a PC-free, measurement-to-actuation layer that converts atom-occupancy readout into optical-tweezer motion commands entirely within FPGA hardware.

Figure~\ref{Fig1}(b) shows the experimental sequence. Each cycle begins
with a $150\,\mathrm{ms}$ magneto-optical-trap (MOT) stage. After light-assisted collisional loading to ensure single-atom occupancy~\cite{schlosserCollisionalBlockadeMicroscopic2002}, single atoms undergo secondary cooling via polarization gradient cooling (PGC) and grey molasses cooling (GMC). The cooling and repumping lasers are then used to perform fluorescence detection on the single atoms. Photons scattered by each atom are coupled into the corresponding OCMs and subsequently transmitted through independently coupled FAs to SPDs, where the optical signals are converted into transistor-transistor logic (TTL) signals. This optical-to-electrical conversion constitutes step  \textcircled{1} in Fig.~\ref{Fig1}(a). Next, the counter card accumulates the photon counts and sends them to the ARM (embedded processor core) processor within it, which compares the collected photon counts against a preset threshold to binarize the trap occupancy into 1 (occupied) and 0 (vacant). (step \textcircled{2} in Fig.~\ref{Fig1}(a)). Based on the atomic positions, the atom rearrangement paths are calculated, and these paths are compiled into radio-frequency (RF) movement commands. These RF commands are subsequently sent to the counter card (corresponding to step \textcircled{3} in Fig.~\ref{Fig1}(a)), which relays them to the arbitrary waveform generator (AWG) card via backplane transmission (corresponding to step \textcircled{4} in Fig.~\ref{Fig1}(a)). Finally, the AWG card outputs the chirped RF waveforms to the AOD for atom movement (corresponding to step \textcircled{5} in Fig.~\ref{Fig1}(a)), completing the rearrangement of the atom array. A subsequent detection step verifies that the target array is defect-free.  

\begin{figure}[hpbt]
\centering
\includegraphics[width=1.0\columnwidth]{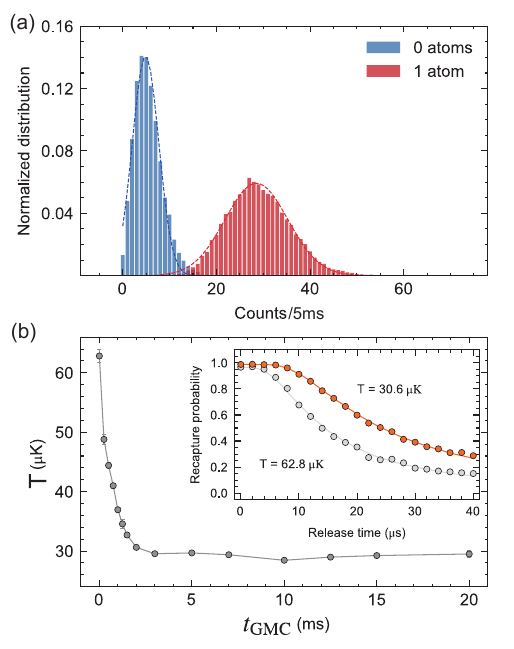}
\caption{\textbf{The normalized histogram distribution of fluorescence counts and the atomic temperature.} (a) The histogram distribution of the channel 6 trap. (b) The atomic temperature varies with the GMC duration, and the saturated atomic temperature is ${30.6}\,\mathrm{\mu K}$ when $t_{\text{GMC}} \ge \text{2} \, \text{ms}$. Inset shows the results of the release and recapture when no GMC (grey points and lines) and {2}\,{ms} GMC (orange points and lines), the atomic temperatures are ${62.8(11)}\,\mathrm{\mu K}$ and ${30.6(3)}\,\mathrm{\mu K}$.}
\label{Fig2}
\end{figure}

\begin{figure*}[hpbt]
\centering\includegraphics[width = 15cm]{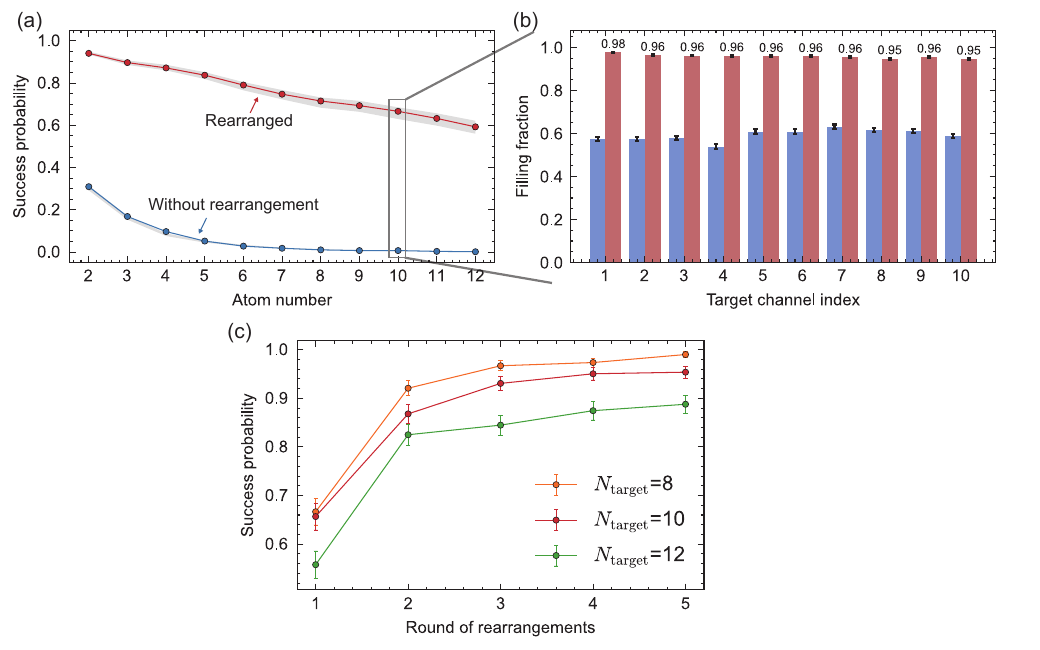}
\caption{\textbf{Results of the single-round and the multi-round rearrangement.} (a) Comparison of the success probability between single-round rearrangement and without rearrangement versus atom number  (b) Comparison of filling fraction between single rearrangement and without rearrangement. (c). Multi-round rearrangement operations effectively enhance the success probability of generating defect-free atom arrays. At round = 5, the success probability is 99.01\% for $N_{\text{target}} = 8$ (orange line), and reaching 95.38\% and 88.78\% for $N_{\text{target}} = 10$ (red line) and 12 (green line).}
\label{Fig3}
\end{figure*}

Figure~\ref{Fig1}(c) illustrates the rearrangement of the one-dimensional
(1D) atom array. In this work we generate a 1D array of 24 optical dipole traps
at $852\,\mathrm{nm}$, with a spacing of $4.5\,\mu\mathrm{m}$ and a trap depth of
$\sim0.8\,\mathrm{mK}$. A vertical line divides the array into two regions: the
target zone (left), where the defect-free array is assembled, and the reservoir
zone (right), which stores surplus atoms for replenishing defects in the target
zone. After the first detection, traps loaded with a single atom are labeled
``1'' (1-atom) and those without ``0'' (0-atom); the 0-atom traps are then
adiabatically switched off. Following the computed movement paths, the ten atoms
nearest the target zone are transported into the target positions, while
reservoir atoms that need no relocation remain in place. Once rearrangement is
complete, the switched-off traps are adiabatically reopened to restore the static
trap configuration. From these runs we extract two quantities: the \emph{success
probability}, defined as the probability that the entire target array is
defect-free over $N_{\mathrm{exp}}$ runs
($N_{\mathrm{target\,full}}/N_{\mathrm{exp}}$), and the \emph{filling fraction}
of the $i$-th tweezer, defined as the probability that a single dipole trap holds
exactly one atom after rearrangement, averaged over $N_{\mathrm{exp}}$ runs
($N_{\text{1-atom},i}/N_{\mathrm{exp}}$).

Among these processes for rearrangement, the total system latency is
\begin{equation}
t_{\text{total}}=t_{\text{read}}+t_{\text{calculate}}+t_{\text{PS}\rightarrow\text{PL}}+t_{\text{PL}\rightarrow\text{PL}},
\end{equation}
where $t_{\text{read}}$ (step \textcircled{2}) denotes the latency for the counter card to read the counts  and transmit them to the ARM processor, $t_{\text{calculate}}$ represents the time for the ARM processor to compute the rearrangement paths, $t_{\text{PS}\rightarrow\text{PL}}$ (step \textcircled{3}) is the latency for transmitting RF commands from the ARM processor system (PS) to the counter card programmable logic (PL) side, and $t_{\text{PL}\rightarrow\text{PL}}$ (step \textcircled{4}) is the latency for transmitting RF commands from the counter card PL side to the AWG card PL side.

Figure~\ref{Fig1}(d) shows the physical implementation of the electronic
control system (PXIe-QC100, Hangzhou Biaozhang Electronics). The chassis adopts PXIe slots and employs a
backplane transmission protocol to achieve high-speed, low-latency signal
transfer. Currently the chassis can accommodate up to eight counter cards and
AWG cards, with each slot integrating a power supply and signal transmission for
the functional cards. The leftmost section houses the PC, which programs the
counting tasks for the FPGA-based counter cards and the RF output for the AWG
cards, coordinating the execution of experimental tasks that require logic
feedback, such as atom-array rearrangement. Once the programs are compiled and uploaded from the host computer, the host computer operates in a fully offline state and no longer participates in any experimental or feedback processes.

\section{Atom array rearrangement}

With the OCM chip, the fluorescence from each trap is guided to individual SPDs, so occupancy detection is intrinsically parallel. Figure~\ref{Fig2}(a) displays, as an example, the histogram of fluorescence counts collected over a $5\,\mathrm{ms}$ detection window for channel 6. The red (blue) data correspond to the presence of 1 (0) atom, and the average readout fidelity $\langle F \rangle$ across the entire array reaches $99.4\%$. During the rearrangement process, the atomic temperature $T$ increases and atom loss may occur, primarily due to the heating from  dipole-beam intensity fluctuations~\cite{zhang2025fast, andersen2022optical, luAstigmatismfree3DOptical2025} and from the acceleration/deceleration of the trap during transport~\cite{cicali2024neutral}. To suppress this loss, we apply GMC~\cite{RN366} during rearrangement, and determine the atomic temperature using the release-and-recapture method~\cite{lettObservationAtomsLaser1988,salomonLaserCoolingCesium1990,zhangOptimizedTemperatureMeasurement2009}. In the absence of GMC (grey points and lines), $T$ reaches ${62.8(11)}\,\mathrm{\mu K}$ after one round of rearrangement. We then applied GMC for varying durations, as shown in Fig.~\ref{Fig2}(b). The temperature decreases continuously with prolonged GMC application, and saturates at a duration of ${2}\,\mathrm{ms}$, at which point the system reaches the GMC cooling limit and further extension becomes unnecessary. The final temperature at this point is $T={30.6(3)}\,\mathrm{\mu K}$ (orange points and lines). Based on these results, we adopt a fixed GMC duration of ${2}\,\mathrm{ms}$ for all rearrangement sequences, which improves the rearrangement success probability by lowering the atomic temperature.

As shown in Fig.~\ref{Fig3}(a), we varied the target array size from $N=2$ to $N=12$ while keeping the total number of initial loading traps fixed at 24. Based on statistics from 300 experimental runs, we extracted the success probabilities for generating defect-free arrays by rearrangement (red data points) versus stochastic loading without rearrangement (blue data points). As the target size increases, the rearrangement success probability decreases approximately linearly with array size, yet remains significantly higher than that obtained by stochastic loading alone; in contrast, the probability of
stochastically obtaining a defect-free array decays exponentially with array size and becomes statistically negligible for $N>8$. The grey shaded region in Fig.~\ref{Fig3}(a) represents the projection based on the mean filling fraction and its standard deviation, evaluated within a $95\%$ confidence interval.

Figure~\ref{Fig3}(b) presents the results of rearranging atoms from a randomly loaded array of 24 optical tweezers into a target zone of $N=10$ atoms. Following the $150\,\mathrm{ms}$ MOT loading step, single atoms are stochastically loaded into the 24 tweezers, giving an initial single-atom filling fraction of approximately $0.6$ in the target region (blue bars). After rearrangement, the single-site filling fraction increases to approximately $0.96$ (red bars), substantially enhancing the probability of generating defect-free arrays; this confirms that, in neutral-atom systems, rearrangement substantially mitigates the limitations imposed by stochastic loading.

In our system, the 24 optical dipole traps load an average of 14 single atoms. For target zone sizes ranging from $N=2$ to $12$, surplus atoms therefore remain available after a single-round rearrangement, which motivates a feedback-based iterative rearrangement protocol that uses these surplus atoms to further improve the success probability of assembling defect-free arrays. Figure~\ref{Fig3}(c) shows the resulting improvement over multiple rearrangement cycles within a single MOT loading cycle: the orange, red, and green curves correspond to target sizes $N_{\text{target}}=8$, $10$, and $12$, respectively. After 5 rounds of rearrangement, the success probabilities improve markedly, from baselines of $66.67\%$, $65.68\%$, and $55.78\%$ to final values of $99.01\%$, $95.38\%$, and $88.78\%$ for $N_{\text{target}}=8$, $10$, and $12$. Because each rearrangement round consumes only surplus atoms, this iterative protocol can, in principle, be extended to actively compensate for atom loss occurring during quantum-circuit execution, provided a reservoir of spare atoms is maintained adjacent to the computational zone. The multi-round rearrangement protocol critically depends on the electronic latency per round; we now characterize each stage of this latency budget.

\section{Characterization of electronic latency}

\begin{figure}[hpbt]
\centering
\includegraphics[width= 0.8\columnwidth]{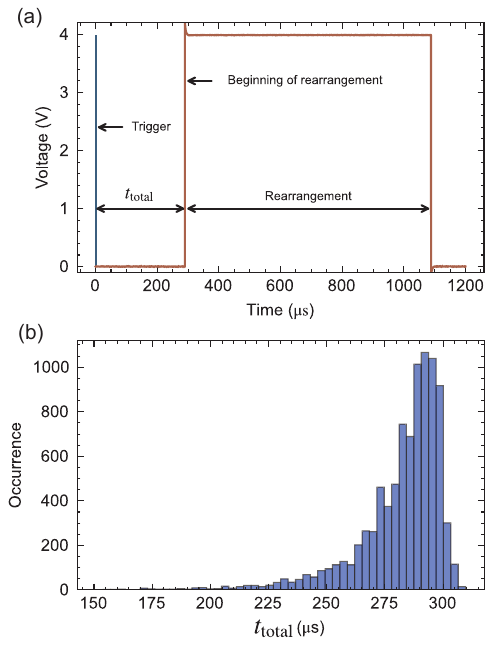}
\caption{\textbf{Results of the total system latency in the rearrangement.} (a) The acquisition of the total system latency $t_{\text{total}}$. (b) The histogram distribution of the total system latency $t_{\text{total}}$ for 8,000 samples.}
\label{Fig4}
\end{figure}

\begin{figure*}[ht]
    \centering
    \includegraphics[width=15cm]{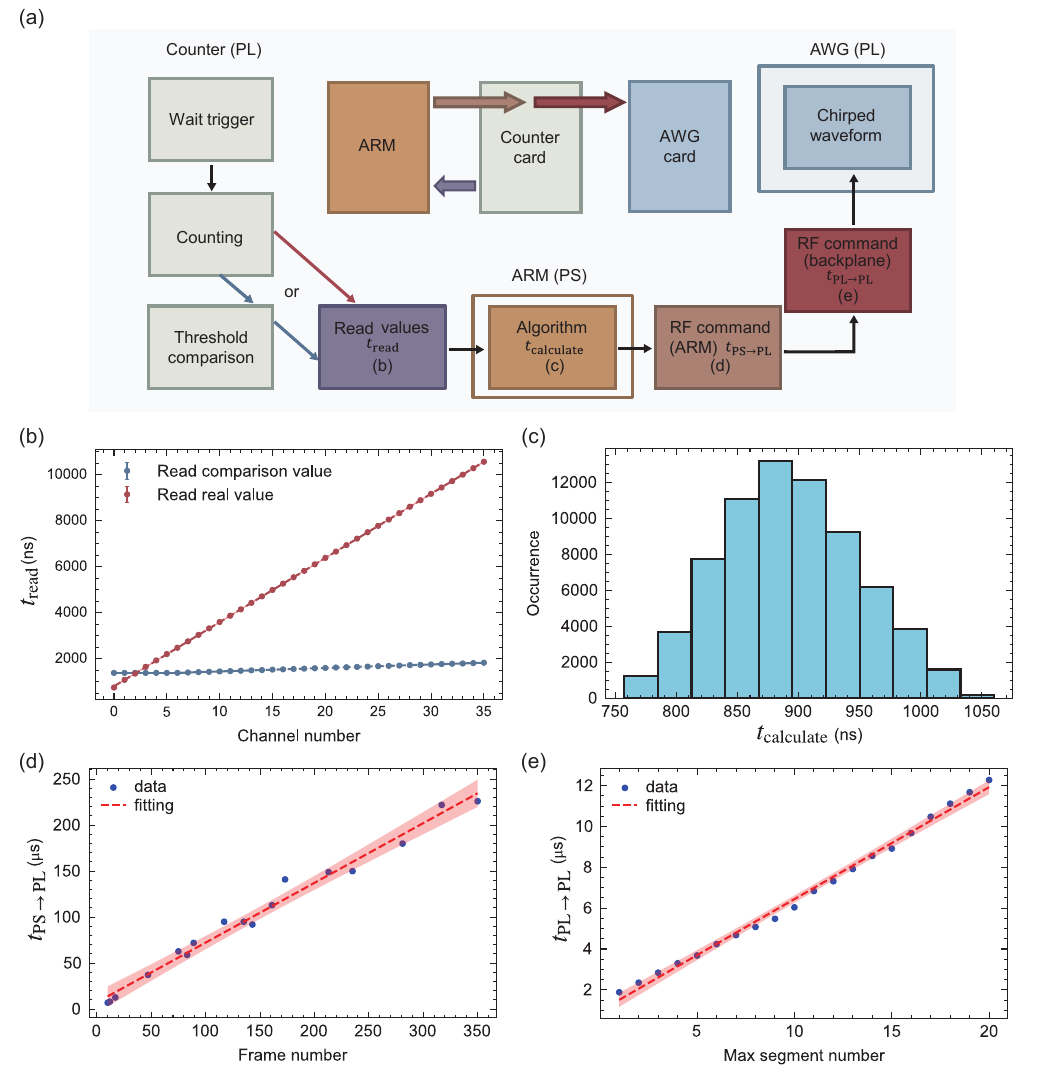}
    \caption{\textbf{Schematic diagram and latency characterization of the electronic control system.} (a) Information flow from atom detection to rearrangement execution, with the total system latency defined as $t_{\text{total}}=t_{\text{read}}+t_{\text{calculate}}+t_{\text{PS}\rightarrow\text{PL}}+t_{\text{PL}\rightarrow\text{PL}}$. (b) Comparison of time overhead between direct real-value readout and comparison-value readout. (c) Histogram of the calculation time for atom transport path assignment in the ARM processor (70,000 samples). (d) Transmission latency from the ARM processing system (PS) to the counter card programmable logic (PL) as a function of frame number. (e) Transmission latency from the counter card PL to the AWG card PL as a function of the longest segment number.}
    \label{Fig5}
\end{figure*}

The PC-free FPGA system enables quick fluorescence collection, efficient path planning, and rapid atom rearrangement. To characterize the total latency of a single-round rearrangement cycle, the AWG card is configured to output a square wave during rearrangement, whose rising edge marks the start of rearrangement and whose falling edge marks its end.
Both this square-wave signal and the detection-completion pulse are fed into a data-acquisition card, with the falling edge of the detection pulse serving as the trigger input. Figure~\ref{Fig4}(a) shows a representative single-shot trace, in which the blue line marks the detection-completion (trigger) pulse and the yellow square wave marks the
rearrangement-indicator signal from the AWG. The total system latency $t_{\text{total}}$ is thus obtained as the time interval between the falling edge of the blue trace and the rising edge of the yellow square wave. Figure~\ref{Fig4}(b) presents the histogram of $t_{\text{total}}$ over 8,000 repeated measurements, yielding $t_{\text{total}}={282(19)}\,\mathrm{\mu s}$ and confirming that the feedback latency is reproducible across runs.

The feedback control path from detection to rearrangement is illustrated in Fig.~\ref{Fig5}(a). To minimize the system latency, our electronic control system employs three optimizations at the transmission level. First, the PC-free electronic system bypasses the additional latency induced by PC communication via the PCIe or USB protocol. Second, the data transmitted to the ARM in the processing system consists of binarized comparison results when detection is completed, rather than the real counts. Third, the rearrangement is performed using simple instructions rather than by directly transmitting the time-domain waveforms.

The counter card initiates the counting task upon receiving a timing trigger. On the PL side, trigger processing, count evaluation, and threshold comparison each consume a single clock cycle (${4}\,\mathrm{ns}$). The readout latency $t_{\text{read}}$ depends on the data format transmitted to the ARM processor, as characterized in Fig.~\ref{Fig5}(b). If the real counts are transmitted, $t_{\text{read}}$ scales linearly with the trap number ($N_{\text{trap}}$) as
\begin{equation}
t_{\text{read}} = 279.01(16)\,N_{\text{trap}} + 809(3)\,\text{ns}.
\end{equation}
By contrast, transmitting only the binarized comparison results yields a substantially lower latency:
\begin{equation}
t_{\mathrm{read}} =
\begin{cases}
{1393.27(29)}\,\mathrm{ns}, & N_{\mathrm{trap}} \leq 7, \\[4pt]
N_{\mathrm{trap}}\times15.14(3)\,\mathrm{ns} + {1305.1(8)}\,\mathrm{ns},
& N_{\mathrm{trap}} > 7 .
\end{cases}
\label{eq:read}
\end{equation}
The results reported here use the real-value mode throughout, since retaining the raw counts simplifies debugging during system characterization, while the comparison-value mode remains available and would reduce $t_{\text{read}}$ accordingly.

Upon receiving the detection results, the ARM processor calculates the rearrangement path. Figure~\ref{Fig5}(c) shows the histogram of the computation time for rearranging atoms from a 24-site array into a 10-site target, yielding
\begin{equation}
t_{\text{calculate}} = {890\pm60}\,\mathrm{ns}
\end{equation}
over 70,000 samples. The rearrangement RF command is then transmitted from the ARM to the counter card PL side. The number of transmitted frames is given by $f = 15N_{\text{move}}+12N_{\text{close}}$, where $N_{\text{move}}$ is the number of atoms to be moved and $N_{\text{close}}$ is the number of traps to be closed. As shown in Fig.~\ref{Fig5}(d), the corresponding transmission latency scales linearly as
\begin{equation}
t_{\text{PS}\rightarrow\text{PL}} = 0.795(17)\,f+4(4)\,\mu\text{s}.
\end{equation}
Subsequently, the RF command is forwarded via the backplane from the counter card PL side to the AWG card PL side. This latency depends on the longest segment number $S$ in the command ($S = 5$--$10$ in our experiments) and follows
\begin{equation}
t_{\text{PL}\rightarrow\text{PL}} = 0.549(10)\,S+0.96(12)\,\mu\text{s},
\end{equation}
as shown in Fig.~\ref{Fig5}(e). By pre-loading compact instructions into the AWG card rather than streaming full-time-domain waveforms, the data transmission latency is further reduced.

\begin{table}[t]
    \centering
    \caption{The time overhead of sub-links.}
    \setlength{\tabcolsep}{10pt}
    \begin{tabular}{lcc}
    \hline\hline
     Sub-links & Parameters & Time overhead \\
     \hline
     $t_\mathrm{read}$  & $N_\mathrm{trap}=24$  & ${7.51}\,\mathrm{\mu s}$ \\
     $t_\mathrm{calculate}$ &-   & $ {0.89}\,\mathrm{\mu s}$\\
     $t_{\mathrm{PS}\rightarrow\mathrm{PL}}$ &$f\approx324$ & $\sim{261.58}\,\mathrm{\mu s}$\\
     $t_{\mathrm{PL}\rightarrow\mathrm{PL}}$ &$S\approx7$ &${4.80}\,\mathrm{\mu s}$\\
     \hline
     $t_\mathrm{total}$ &-  & $\sim{274.78}\,\mathrm{\mu s}$\\
     \hline\hline
    \end{tabular}
    \label{Table1}
\end{table}

\section{Discussions and conclusions}

This work presents a low-latency electronic control system (PXIe-QC100) that removes the communication-bandwidth bottleneck of conventional PC-based control systems, reducing the total feedback latency to ${282(19)}\,\mathrm{\mu s}$. Built on this system, we demonstrate parallel atom-presence detection using a multichannel counter card together with an optical channel mapping device, achieving a $96\%$ filling fraction in single-round rearrangement and, through multi-round rearrangement, a $95\%$ success rate for assembling 10 atoms from 24 tweezers. This remaining single-round limit is set by atomic physics rather than by the control electronics: a lower atomic temperature and a smoother transport trajectory should raise the filling fraction further. The PXIe chassis used here already scales beyond what is demonstrated, from the 8 slots used in this work to 18 within a single chassis, with synchronization available across multiple chassis for still larger arrays.

The dominant remaining latency cost is architectural rather than physical: the ARM processor is currently embedded in the counter card, a placement chosen to support 128 sub-waveforms on the AWG card, so every rearrangement instruction is routed through an extra counter-card transmission step that dominates the latency budget. Moving the ARM into a higher-performance AWG-card FPGA instead would remove this step without sacrificing sub-waveform capacity. Under this redesign, we project a  readout latency of $600\,\mathrm{ns}$ (real values) or ${37.5}\,\mathrm{ns}$ (binary values) per channel, and a corresponding counter-card-to-ARM transmission overhead of ${800}\,\mathrm{ns}$ or ${25}\,\mathrm{ns}$, a path-calculation time of about $1\,\mathrm{\mu s}$, and an ARM-to-AWG overhead of $340\,\mathrm{ns}$ per sub-waveform. Scaling these projected figures to larger arrays gives a total latency below ${35}\,\mathrm{\mu s}$ for 64 atoms and about ${400}\,\mathrm{\mu s}$ for 1000 atoms. Meanwhile, the same measurement setup applies directly to the internal state (e.g., hyperfine state) discrimination. Combining with mid-circuit measurement~\cite{grahamMidcircuitMeasurementsSingleSpecies2023,ma2023highfidelitya}, this would in principle allow real-time extraction of error syndrome measurement and execution of dynamic quantum circuits~\cite{Corcoles2021}.

\begin{acknowledgements}
This work was funded by the National Key R\&D Program (Grant Nos.~2021YFA1402004,~2021YFA1402002 and 2025YFF0515200), and the National Natural Science Foundation of China (Grants Nos.~92465201 and 92265210). Y.-L. Z. was supported by the Shenzhen International Quantum Academy (Grant No.~SIQA2025KFKT02). This work was supported by the Fundamental Research Funds for the Central Universities and USTC Research Funds of the Double First-Class Initiative. This research was also supported by the Supercomputing Center of USTC and the USTC Center for Micro and Nanoscale Research and Fabrication.
\end{acknowledgements}

%

\end{document}